\documentclass[doublecol]{epl2}

\usepackage{epsfig}
\usepackage{color}
\usepackage{amsmath}
\usepackage{amssymb}
\usepackage{graphicx}

\title{String structures in driven 3D complex plasma clusters}
\shorttitle{}

\author {L. W\"{o}rner\inst{1,}\inst{2}, C. R\"{a}th\inst{1}, V. Nosenko\inst{1}, S. K. Zhdanov\inst{1}, H. M. Thomas\inst{1}, G. E. Morfill\inst{1} \\ \\ J. Schablinski\inst{3}, D. Block\inst{3}}

\institute{
\inst{1} Max Planck Institute for Extraterrestrial Physics, Giessenbachstrasse, 85748 Garching, Germany\\
\inst{2} Groupe de la Recerche sur l'Energie et les Materiaux Ionis\'ees, F-84677 Orl\'eans, Cedex 2, France\\
\inst{3} Christian-Albrechts Universit\"at zu Kiel, D-24118 Kiel, Germany}

\shortauthor{}


\abstract{
The structure of driven three-dimensional complex plasma clusters was studied experimentally. The clusters consisted of around 60 glass microspheres that were suspended in a plasma of rf discharge in argon. The particles were confined in a glass box with conductive yet transparent coating on its four side walls. This allowed manipulating the particle cluster by biasing the confining walls in a certain sequence and direct imaging of the cluster. In this work, a rotating electric field was used to drive the clusters. Depending on the field frequency, the clusters rotated ($10^4-10^7$ times slower than the rotating field) or remained stationary. The cluster structure was neither that of nested spherical shells nor simple chain structure. Strings of various lengths were found consisting of 2 to 5 particles, their spatial and temporal correlations were studied. The results are compared to recent simulations.
}

\begin{document}

\maketitle

\section{Introduction}
\label{sec:intro}


A complex (dusty) plasma consists of a weakly ionized gas and small solid particles \cite{Morfill:2009}. The particle size can be from tens of nanometers to tens of microns. The particles are usually negatively charged due to the high thermal speed of plasma electrons. In ground-based experiments, the particles can be levitated against gravity by a strong electric field present in the plasma sheath (pre-sheath) around an electrode or wall \cite{Thomas:1994,Nosenko:2012}. Horizontally, like-charged particles are held together by weak electric fields naturally present in the plasma or by external confinement. The latter can be provided, e.g. by a recession in the electrode \cite{Konopka:2000} or by additional containers such as a glass box \cite{Arp:2004, Ivanov:2009, Nosenko:2009}.

The interparticle interaction in a suspended particle cloud consists of two parts: direct screened-Coulomb (Yukawa) repulsion and ion-wake mediated attraction. The ion wakes are formed by ions streaming past a micro-particle due to their focusing behind it \cite{Lampe:2000, Miloch:2010, Kroll:2010:1, Ivlev:2008}.

The structure of a suspended particle cloud is therefore determined by the interplay of the isotropic Yukawa-type and anisotropic wake-field forces giving rise to different competing symmetries and patterns, e.g. nested spherical shells \cite{Arp:2004} and flow-aligned strings \cite{Melzer:2001, Melzer:2010}. The order transition between these two configurations was found to be governed by the ion flow Mach number in the dust dynamics simulations \cite{Ludwig:2012}.

Structuring is one of the basic processes in complex plasmas, as well as many other systems. For instance, dust chains, filaments, and stationary structures have been observed in sparks \cite{Marsh:1910} and dc plasmas \cite{Molotkov:1999,Merlino:2011}. Note that competing repulsive-attractive interactions often define the actual variety of observed strings, filaments and patterns (lattices) formed by the interacting strings as, for instance, in colloids \cite{Reichhardt:2004}, in superconductors \cite{Silhanek:2010}, in electrons on superfluid helium \cite{Rousseau:2009}, and in current-carrying plasmas \cite{Trubnikov:2002}.

To probe the structure of clusters in various states, some method of external manipulation is necessary. One way of manipulating a cluster is applying a torque to it. This can be done by means of external magnetic field \cite{Konopka:2000}, electric field \cite{Nosenko:2009, Woerner:2011}, or by the drag force of rotating neutral gas \cite{Kroll:2010:1}.

In the present paper, we studied a particular class of \emph{dynamically driven clusters} (DDC) -- 3D clusters of charged particles driven by rotating electric field. The presence of the driving force deeply affected the actual cluster geometry and the global symmetry of the cluster that appeared to be defined by spontaneously formed particle strings interacting with each other. The recognition of string structures observed in DDC's is the main goal of this study.

The lattices as well as competing symmetries shown by the string structures formed in our experiments resemble to some extent the Faraday wave patterns parametrically excited in water by vertical vibrations \cite{Douady:1990}. Similarities include, for instance, the competition between sub-harmonically generated equilateral triangles and regular hexagons  \cite{Kumar:1994} and the super-lattice patterns \cite{Fineberg:2002}. Therefore, experiments with complex plasma clusters might help to get insight into generic structuring processes at the kinetic level.

\section{Experimental setup}
\label{setup}

Our experimental setup has been previously described in Ref.~\cite{Woerner:2011}. We used a capacitively coupled rf discharge in argon at a pressure of $4$~Pa. The inner dimensions of the chamber are $14~{\rm cm} \times 14~{\rm cm} \times 26~{\rm cm}$ \cite{Kroll:2008}. The lower electrode diameter is $7$~cm. The upper electrode is mounted at a distance of $7.2$~cm and has a diameter of $9.6$~cm. The lower electrode is powered by a rf generator operating at $13.56$~MHz and $120$~V peak-to-peak amplitude. The upper electrode and the surrounding walls are grounded. In the center of the upper electrode a grid-covered hole is made through which particles can be transferred into the chamber. The particles used in these experiments are hollow glass spheres \footnote{``Glashohlkugeln'' or ``glass bubbles'' produced by 3M GmbH Deutschland, Industriestra\ss e, 55743 Idar-Oberstein.} with a diameter of $22\pm2~\mu$m.

On top of the lower rf electrode we placed a glass box as described in Refs.~\cite{Nosenko:2009, Woerner:2011}. The inner surfaces of the box side plates are coated with Indium Tin Oxide (ITO) and are therefore conductive yet transparent. The plates are insulated from each other and from the lower electrode by plastic poles. Each plate is electrically connected to a separate function generator. The function generators output sinusoidal signals with a peak-to-peak amplitude of $20$~V and frequency in the range of $100$~Hz to $100$~kHz. It is possible to make the resulting electric field inside the box to rotate in the horizontal plane by setting the phase shift of the sinusoidal voltage on the $n$-th plate to $\pm n\pi/2$. This ``rotating wall'' technique was introduced in Ref.~\cite{Nosenko:2009} to drive 2D clusters of particles.

To create a cluster, microparticles were suspended inside the glass box where they formed a spheroidal cloud with a diameter of $\sim 5$~mm. Applied rotating electric field contracted the cluster and forced it to rotate (in fact, co-rotate in the present experiments). The speed of rotation depended on the applied field frequency \cite{Nosenko:2009, Woerner:2011}. Cluster rotation was the fastest at $5$~kHz and practically absent at $1$~kHz \cite{Woerner:2011}, therefore in this paper we used these two frequency settings to study rotating and stationary clusters.

\begin{figure}
\centering
\includegraphics[width=0.95\linewidth]{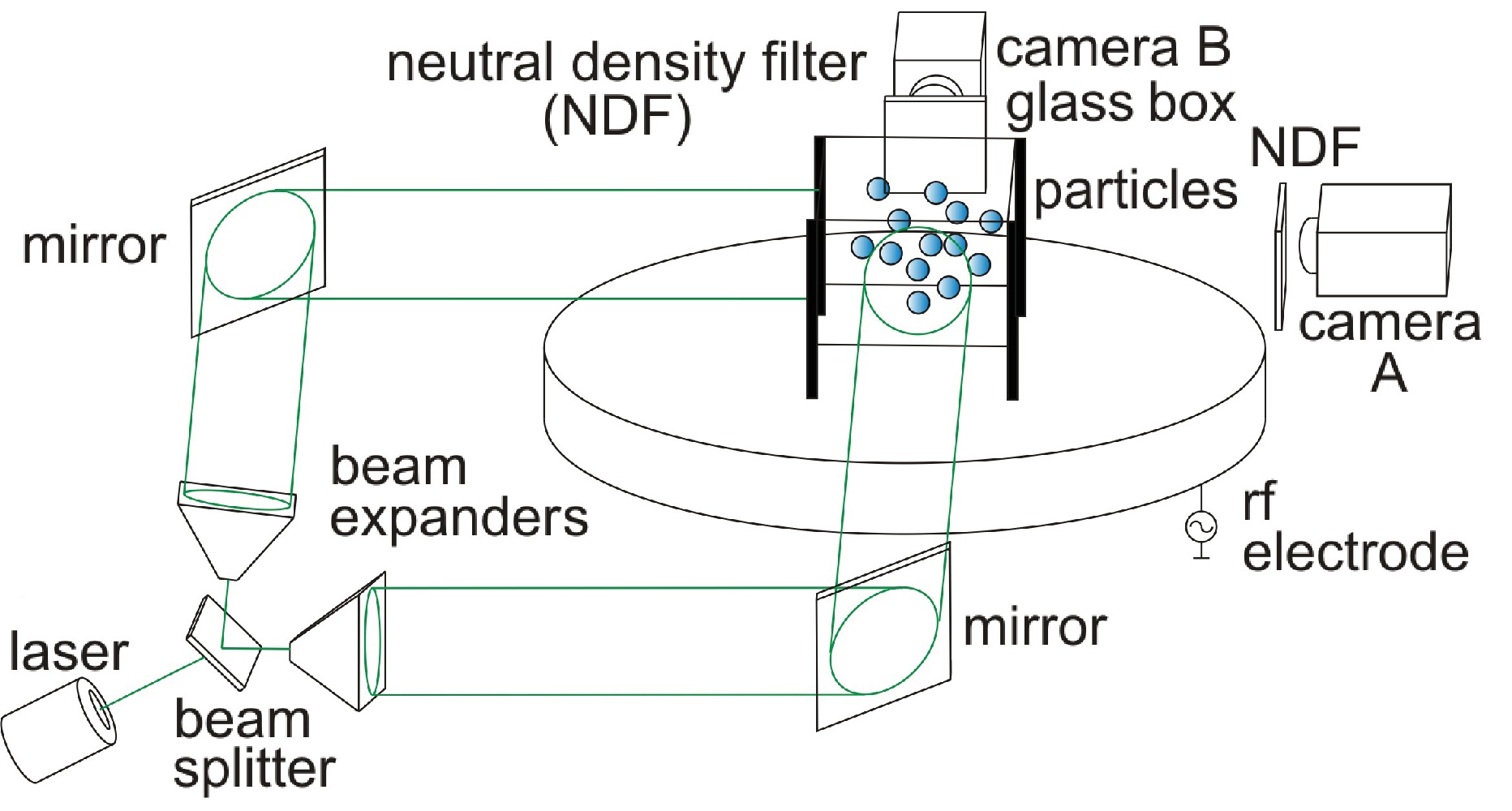}%
\caption{Sketch of the experimental setup. The microspheres are confined in a glass box placed on the lower electrode in argon rf discharge. 3D diagnostic system \cite{Kroll:2008} consists of a laser (wavelength of $532$~nm), beam splitter, and two identical channels consisting of a beam expander, mirror, neutral density filter, and digital camera. Interference patterns created by particles are registered directly by camera sensors. The resulting images contain complete 3D information on particle positions.}
\label{Expset}
\end{figure}

To image the particles, we used a 3D diagnostic method that is called stereoscopic digital in-line holography \cite{Kroll:2008}. This method provides the information on 3D fully-resolved particle dynamics. A sketch of the laser and camera setup is shown in Fig.~\ref{Expset}. The cameras operated at a frame rate of $50$ frames per second during a time interval of $10$~s. No lenses were mounted on the cameras. The laser light was therefore registered directly by the camera sensors, after passing through the particle cloud. In the resulting images, each particle was represented by a system of concentric interference rings, the depth information was encoded in the inter-ring spacing.

\section{Particle cloud parameters}
\label{parameters}

The hollow glass spheres used in our experiments are made of Scotchlite S22 glass with a density of $2.5~{\rm g/cm}^3$ and have a wall thickness of $0.3~\mu$m. They are sorted to achieve a narrow distribution of particle diameters of $22\pm2~\mu$m. It is straightforward to estimate the particle mass $m=1.11\pm0.23$~ng and the neutral gas damping rate\footnote{When calculating $\gamma_{\rm eff}$, one has to take into account that the particles are hollow, unlike discussed in Ref.~\cite{Epstein:1924}.} $\gamma_{\rm eff}=17.5~{\rm s}^{-1}$ (at the argon pressure of $4$~Pa).

In our experimental conditions, the plasma parameters are estimated as follows\footnote{Typical bulk plasma parameters, see Ref.~\cite{Woerner:2011}.}: the electron temperature is $T_e\approx3$~eV, electron-to-ion temperature ratio $T_e/T_i \approx 100$, plasma density $n \approx 10^9~{\rm cm}^{-3}$. The electron Debye screening length is estimated as $\lambda_{De}\approx 400~\mu$m and the ion mean free path as $\lambda_{ia}\approx 770~\mu$m \footnote{Ion-atom charge exchange collisions \cite{Robertson:2003} are assumed to be dominant.}.

The particle charge number is expected to be in the range of $Z=40\,000-50\,000$, as standard theoretical approximations allowed us to estimate (for instance, $Z\simeq 46\,000$ follows from OML \cite{Bonitz:2010}, $Z\simeq 39\,000$ from DML \cite{Morfill:2006}, or $Z\simeq 41\,000$ from modified OML \cite{Khrapak:2005}). Note that in our experimental conditions the ratio $N_D/Z\sim$~2-3 is not large, where $N_D$ is the average number of electrons inside the Debye sphere. Therefore, the actual particle charge is expected to be smaller due to electron depletion \cite{Allahyarov:1998}.

In experiments with clusters of only $2-3$ charged particles (arranged as a string), the particle charge and damping rate can be measured directly by exploring the response of a cluster to harmonic excitation applied to the lower electrode \cite{Carstensen:2012}. A series of such measurements showed, for instance, that in doublets the upper particle charge number was about $Z=40\,000$, while the lower particle's charge was up to $30\%$ smaller. The effective damping rate ($\sim 10~{\rm s}^{-1}$) was measured to be in a reasonably good agreement with the above predictions.

Given the approximate values of particle charge and mass and assuming that the simple balance equation $mg=ZeE$ is valid,
where $g$ is the acceleration of gravity, $e$ is the elementary charge, and $E$ is the local electric field at the particle levitation height, one can estimate the Mach number\footnote{Here, the ratio $M=u_i/c_{si}$ of the ion flow velocity $u_i$ to the speed of ion sound $c_{si}=\sqrt{T_e/m_i}$, where $m_i$ is the ion mass.} of the ion flow. In the mobility-limited regime the following relationship can be used:
\begin{equation}
\label{eq_4}
M=\sqrt{\frac{mg\lambda_{ia}}{ZT_e}}.
\end{equation}
It yields $M=0.6-0.7$ in our experimental conditions. This finding is important for the analysis of cluster structure in the next section, since it allows us to compare our experimental results to the dust dynamics simulation reported in Ref.~\cite{Ludwig:2012}, where $M$ was found to govern an order transition in 3D clusters.

\begin{figure}
\centering
\includegraphics[width=1.0\linewidth]{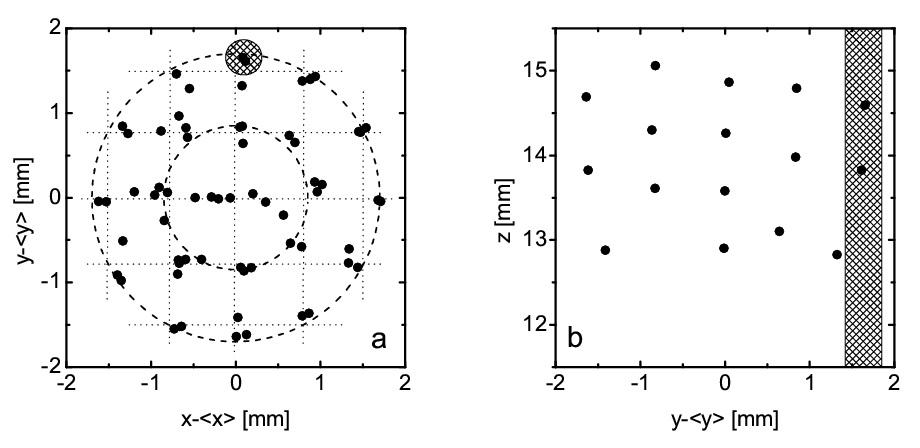}
\includegraphics[width=1.0\linewidth]{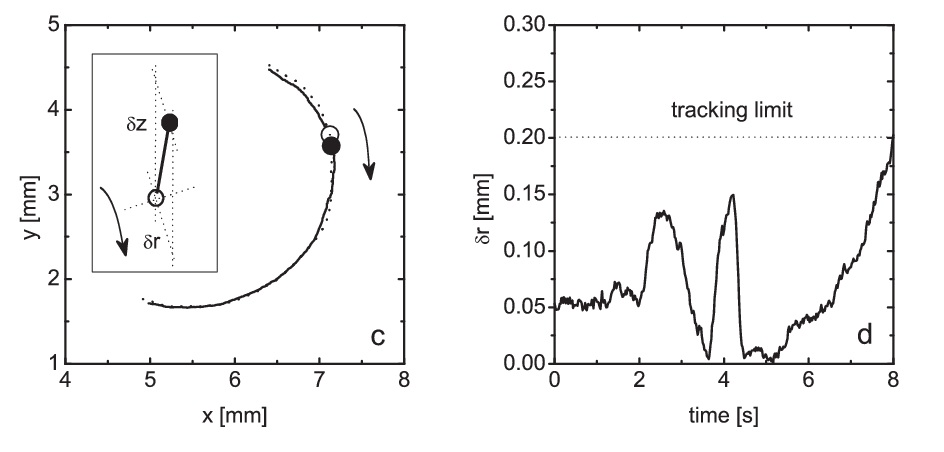}
\caption{String structure of a rotating 3D cluster.  (a) In the top view, the ($x,y$) particle positions (black dots) are structured either as a system of cylindrical shells (dashed lines) or tetragons (dotted lines). The inner dashed circle has a radius of $0.85$~mm, the outer of $1.7$~mm. The grid (dotted lines) has a box side length of $0.77$~mm. The dotted and dashed lines are shown to guide the eye. (b) In the side-view cross-section (defined by $|x-\left<x \right>|\leq0.3$~mm), the vertical strings of particles are clearly seen. The shadowed circle in (a) and shadowed rectangle in (b) indicate the position of an identified two-particle string. The trajectories of these particles during $10$~s are shown in (c). The solid circle and solid line (open circle and dashed line) indicate the positions of the upper (lower) particle in the doublet. The doublet rotates clockwise as sketched in the inset.  (d) Relative horizontal distance between the doublet components as a function of time. The dotted line indicates the tracking limit $\delta r\leq0.2$~mm. The $64$-particle cluster was driven by a clockwise rotating electric field with a frequency of $5$~kHz.}
\label{structure}
\end{figure}

\section{Competing symmetries in driven clusters}
\label{symmetry}

To identify string structures in particle clusters, we analyzed the spatial and temporal correlations in particle positions\footnote{The details of computational procedure will be published elsewhere.}.

\subsection{Spatial correlation}
\label{subsection:spatial}
Horizontal distances between all particles in the $(x,y)$ projection (top view) were calculated, resulting in a distance matrix. Particles were considered as coupled in a string if the horizontal distance between them was below the threshold introduced by the ``critical cylinder'' (oriented vertically) of radius $r_{\rm cyl}=200~\mu{\rm m}\approx\frac12\lambda_{De}$, as illustrated schematically in Figs.~\ref{structure}(a),(b). (As it will become clear later, the chosen value of $r_{\rm cyl}$ is smaller than the mean interstring distance in the $(x,y)$ plane.)

The top view of the cluster in Fig.~\ref{structure}(a) indicates that the majority of particles are organized in vertical strings. The side-view cross-section in Fig.~\ref{structure}(b) also clearly shows particle strings.  Only few particles remain alone. Furthermore, the spacing between two strings is larger than the characteristic horizontal extent of a string. Therefore, most particles are well aligned in the vertical direction (along the $z$-axis).

\subsection{Temporal correlation}
\label{subsection:temporal}
Another important criterion is the lifetime of a particle string. Studying the lifetime of particle associations one can easily distinguish pairs of particles which only occasionally happened to be close to each other from \textit{persistent} particle pairs or \textit{strings} where particles interact with each other by the wake-field attraction. Two or more particles can be assumed to form a string if they are fairly close to each other for a considerable amount of time that is longer than the ``fly-by'' time \footnote{The fly-by time is $\propto \Delta/v$, where $\Delta$ is the interparticle separation and $v$ is a characteristic particle speed.}.

\begin{figure}
\centering
\includegraphics[width=1.0\linewidth]{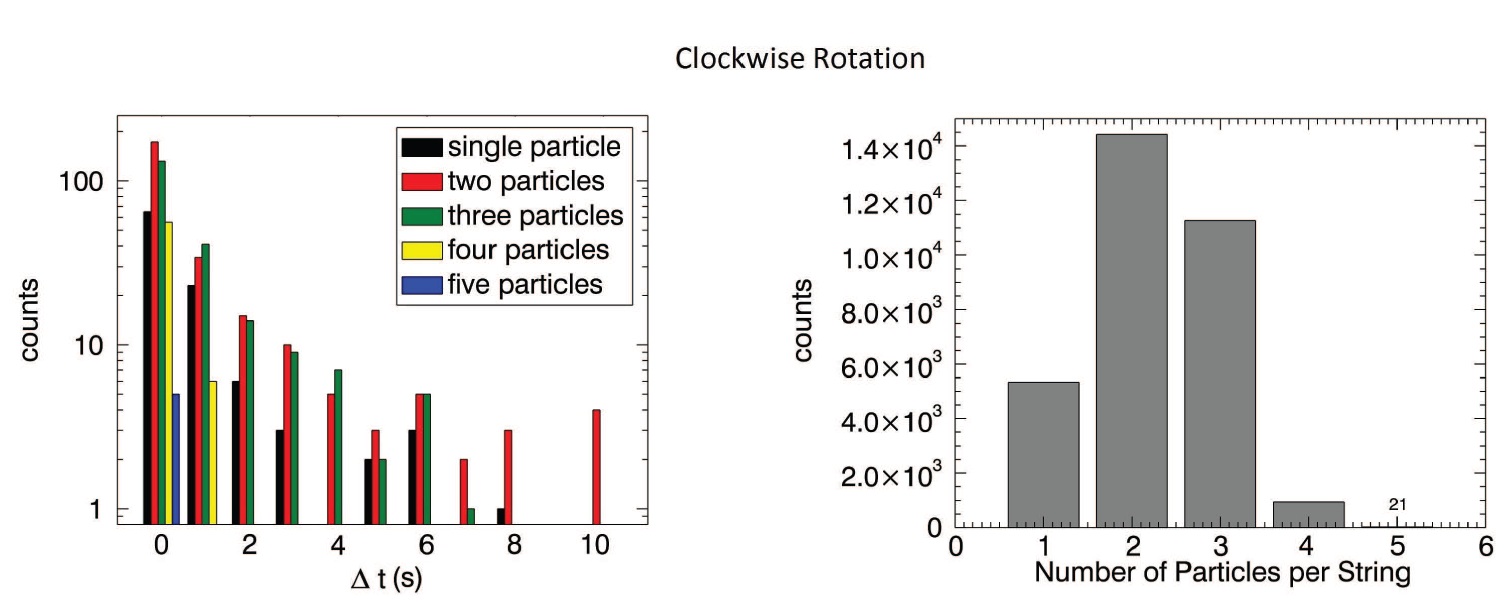}
\includegraphics[width=1.0\linewidth]{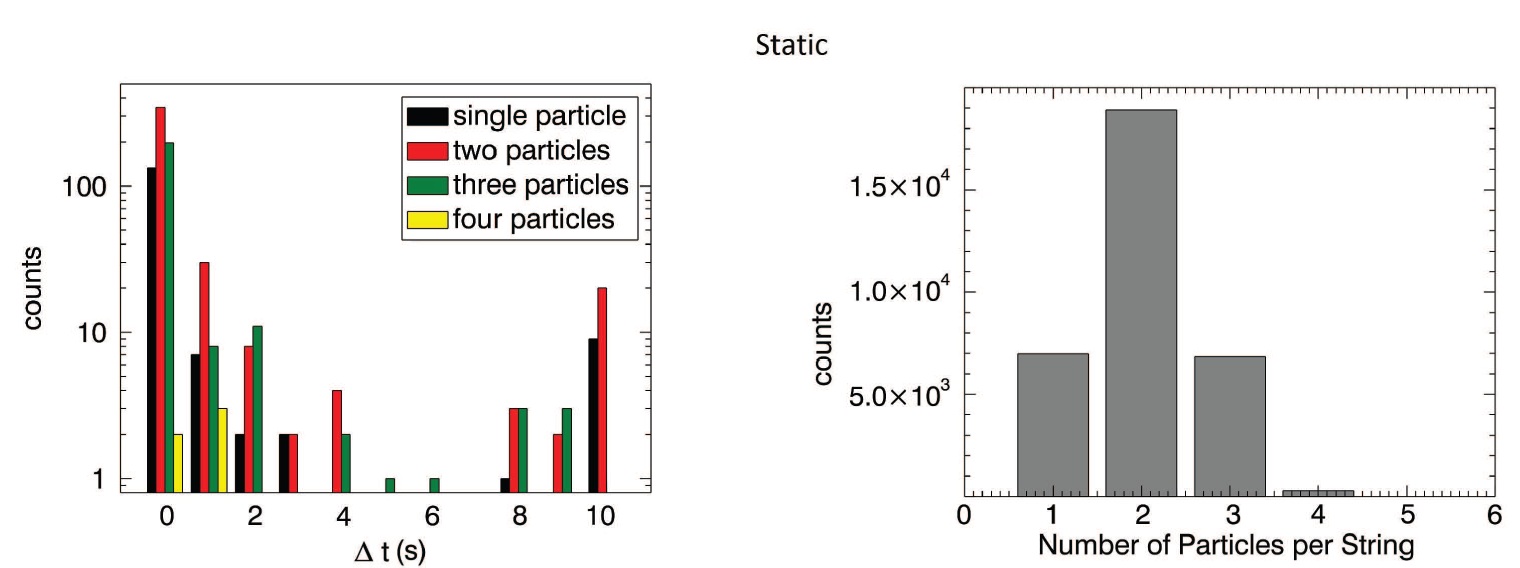}
\caption{(left panels) Histograms of the string lifetimes $\Delta t$ shown separately for strings composed of 2 to 5 particles, as well as single particles. (right panels) Histograms of the number of particles per string. Doublets and triplets are the most probable compositions of strings in a rotating cluster. The experiment was performed with a $64$-particle cluster driven by a rotating electric field with a frequency of $5$~kHz (rotating cluster, top panels) and $1$~kHz (stationary cluster, bottom panels).}
\label{lifetimelog}
\end{figure}

The left panels in Fig.~\ref{lifetimelog} show the histograms of the lifetimes of various string compositions observed in our experiments. The single particle lifetimes are also shown to emphasize the role of the interstring association and dissociation processes. The right panels in Fig.~\ref{lifetimelog} show the histograms of the number of particles per string. It can be seen that doublet string is the most probable state for string configuration, and the most long-living as well.

The observed string structure of 3D clusters agrees with the dust dynamics simulation of Ref.~\cite{Ludwig:2012}. In the simulation, the clusters had a structure of nested spherical shells at $M=0$, a state where vertically aligned pairs of particles form and move through a liquid complex plasma at $M\geq0.16$, a similar ``triplet state'' at $M\geq0.33$, and finally well aligned strings of particles at $M=1$. Our experiment with estimated $M=0.6-0.7$ fits well the above classification.

\subsection{String lattices}
\label{subsection:lattices}
Driven clusters in our experiments possessed large-scale symmetry. The observed cylindrical shell structure had either circular symmetry [indicated by dashed circles in Fig.~\ref{structure}(a)] or tetragonal lattice symmetry [indicated by dotted lines]. Both states are metastable and the actual symmetry of the cluster changed from one to the other while the cluster rotated -- a behavior also found in colloids \cite{Reichhardt:2004} and in quantum states in superconductors \cite{Silhanek:2010}.

\subsection{String length}
\label{subsection:length}

\begin{figure}
\centering
\includegraphics[width=1.0\linewidth]{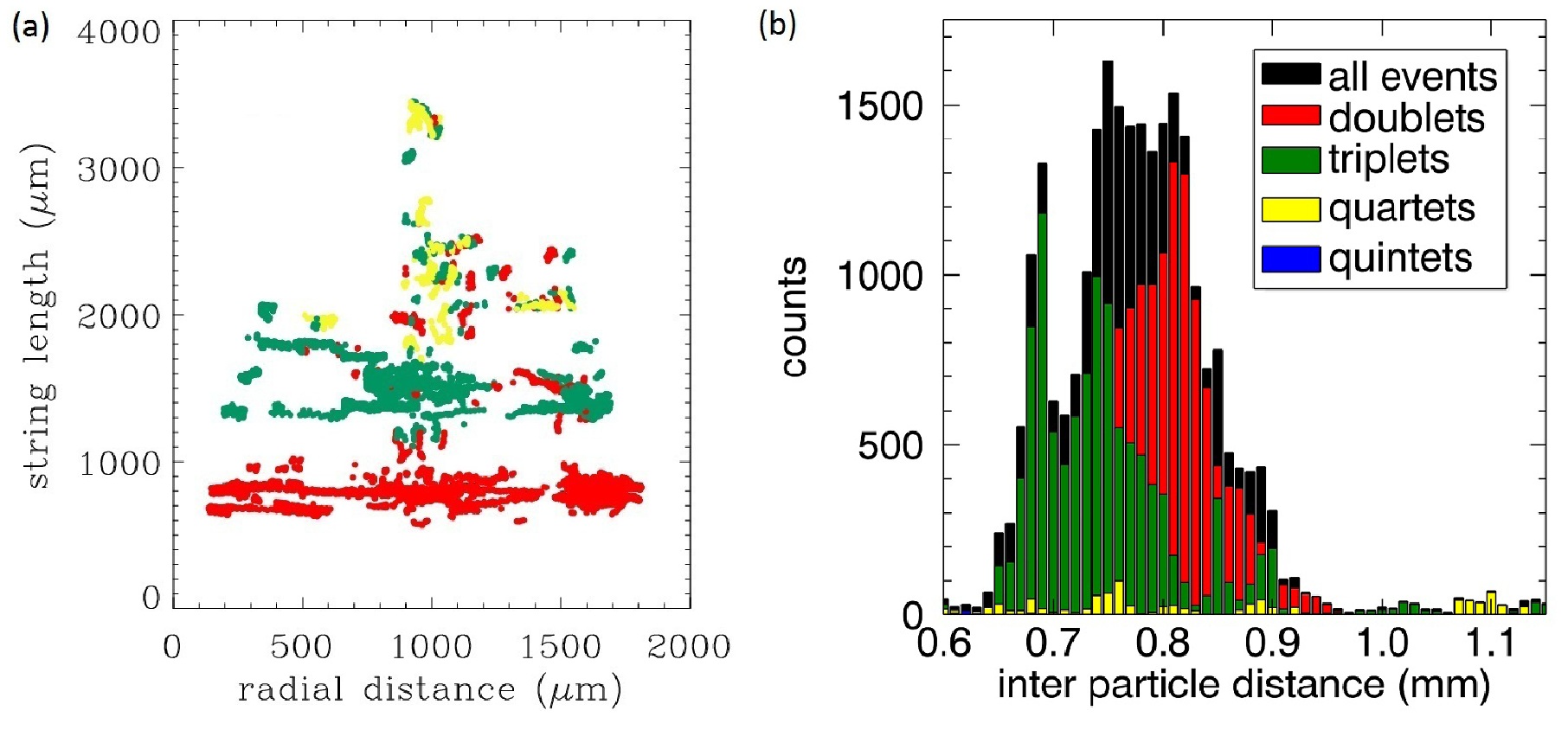}
\caption{String length and in-string interparticle distance. (a) Radial distribution of the string length. The cluster center of mass is located at the origin. (b) Histogram of the in-string interparticle distance. The string composition is color coded as follows: (black) all events, (red) doublets, (green) triplets, (yellow) quartets, (blue) quintets. The experiment was performed with a 64-particle clockwise rotating cluster driven by a rotating electric field with a frequency of $5$~kHz.
}%
\label{rtilt}%
\end{figure}

The spatial distribution of strings in a DDC is not uniform, see Fig.~\ref{rtilt}(a). Two main effects can be seen immediately: (i) the discreteness of the string length distribution is well pronounced and (ii) surprisingly, longer strings are located predominantly not in the middle of the cluster. Another surprising fact is that on average the triplets are relatively more compact than doublets, see Fig.~\ref{rtilt}(b). The overall length of a string composed, e.g. of three particles may be longer than that composed of only two particles. However, on average the three-particle strings are internally more compact, and this is a rather noticeable effect; the string shrinkage is $\sim30$\%. (Note that 2D plasma crystals are nearly incompressible in the elasticity sense \cite{Zhdanov:2012}.) The average particle separation for doublets is $0.87\pm 0.10$~mm, while for triplets it is $0.78\pm 0.10$~mm. For longer strings the statistics is poor. Taking into account all events shown in Fig.~\ref{rtilt}(b), a mean interparticle distance in the vertical direction $\Delta_{\rm vert}=0.83\pm 0.10$~mm can be estimated.

As a whole, the cluster consists of an outer spheroidal shell enclosing a cylindrically-symmetric network of mutually interacting strings. The globally spheroidal shape of the DDC's is most probably determined by the external confinement.

\section{Particle dragging}
\label{subsec:drag}

\begin{figure}[t]
\centering
\includegraphics[width=1.0\linewidth]{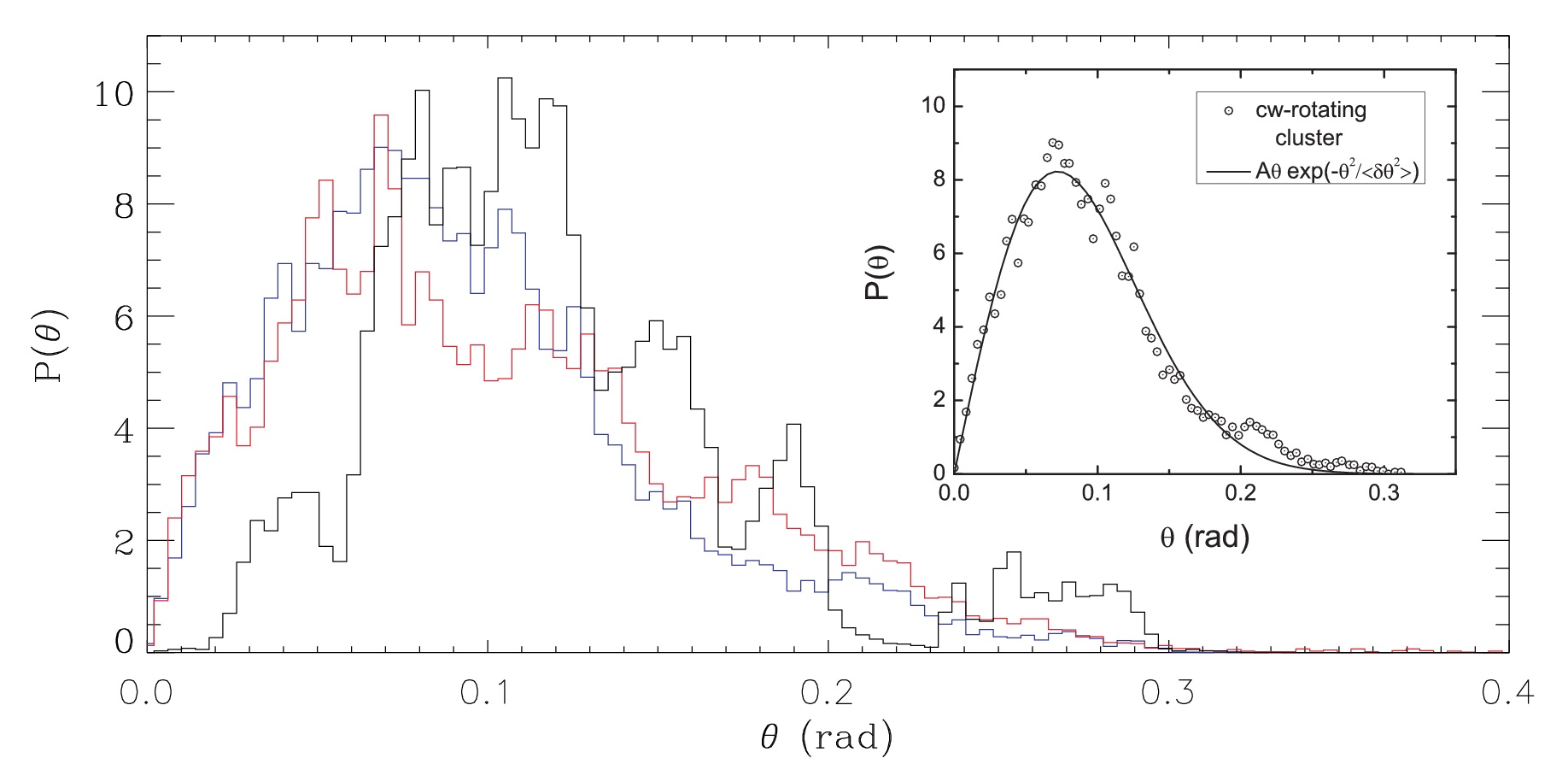}
\includegraphics[width=1.0\linewidth]{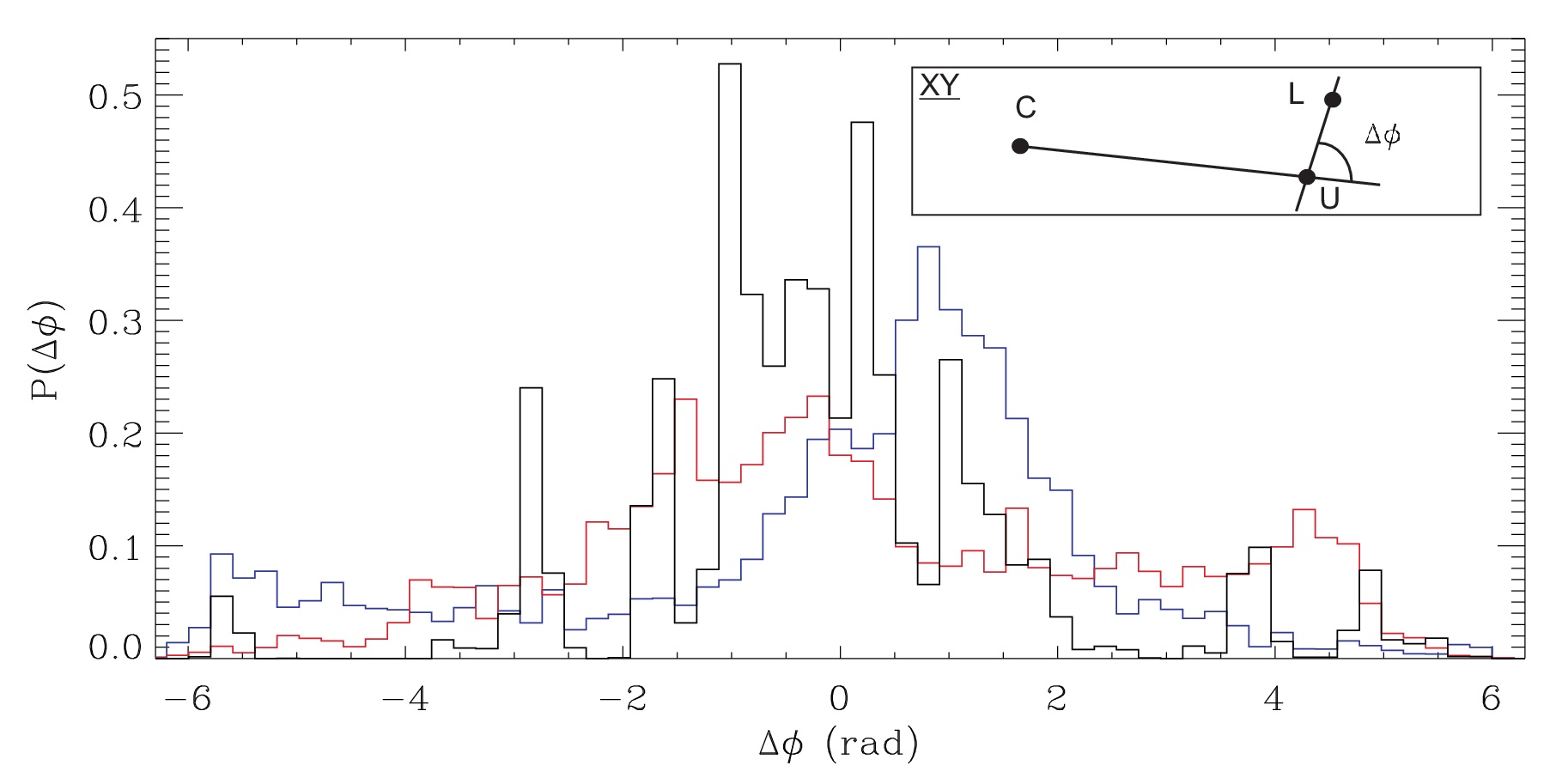}
\caption{Probability densities of (top) polar $\theta$ and (bottom) azimuthal $\Delta\varphi$ angles for particle strings in a 3D cluster. The probability densities are calculated from the respective distributions by normalizing the area under the curve. They are shown for three situations: counterclockwise rotation (red line, rotating electric field at $5$~kHz), clockwise rotation (blue line, $5$~kHz), and no rotation (black line, $1$~kHz). The angle $\theta$ was measured between the vertical $z$-axis and the (imaginary) line connecting the upper and lower particles of a string. (top inset) $P(\theta)$ is fitted well with Eq.~\ref{eq_7}, where $A=188.2\pm3.7$, $\left<\delta \theta^2\right>=0.0104\pm0.0002$. (bottom inset) $\Delta\varphi$ is defined as the relative angular position on the $(x,y)$ plane of the upper (U) and lower (L) particles of a string. C is the cluster center of mass. The range of $\Delta\varphi$ was extended to $[-2\pi,2\pi]$ to take into account two turns made by a particle string.}
\label{angles}
\end{figure}

We observed that in a particle pair in a rotating cluster, the lower particle lags behind the upper one, or is dragged by it, as shown in Fig.~\ref{structure}(c). A convenient way of studying this effect is to plot the distribution of azimuthal angles $\Delta\varphi$ of particle pairs, see Fig.~\ref{angles} (bottom panel). In the stationary cluster (black line), $\Delta\varphi$ has a broad distribution roughly centered at zero. In the counterclockwise rotating cluster (red line), an additional peak develops at $\approx -\pi/2$. In the clockwise rotating cluster (blue line), the peak shifts to $\approx \pi/4$. These peaks at $\Delta\varphi \neq 0$ indicate the particle dragging effect. The reason why the peak in the latter case appears at $\approx \pi/4$ instead of the expected value of $\pi/2$ is not clear.

Observed particle dragging is an important effect, which directly demonstrates the presence of in-string \emph{attraction}\footnote{To some extent, this resembles the laser-dragging experiment of Ref.~\cite{Melzer:2001}.}. In our experiments, the driving torque on the cluster depends on the vertical coordinate \cite{Woerner:2011}. A dragging is then to generate an effective coupling mechanism in response to the driving torque inhomogeneity. In other words, the transverse component of the attraction force to the ion focus $F_w$ is to compensate for the moment deficit $\Delta \mathcal{M}$. $F_w$ evidently could not exceed the friction force $F_{\rm fr}$, that is:
\begin{equation}
\label{eq_5}
F_w=\frac{\Delta \mathcal{M}}{R}\leq F_{\rm fr}=m\gamma_{\rm eff}\omega_cR,
\end{equation}
where $R$ is the distance to the cluster center, $\omega_c$ is the cluster rotation speed. The relationship (\ref{eq_5}) yields an upper estimate $F_w<10$~fN for our parameters\footnote{To compare, the force scale of Yukawa-interacting particles in a sting is about $200-300$~fN, the energy scale is about $300-500$~eV.}.

It is natural to introduce a ``spring constant'' $k_w$ of this wake-mediated interaction:
\begin{equation}
\label{eq_6}
k_w=\frac{F_w}{\left<\delta r\right>},
\end{equation}
where $\left<\delta r\right>$ is the mean relative radial displacement of the dragged particle. For instance, for the doublet traced in a rotating cluster shown in Fig.~\ref{structure}(c),(d) $\left<\delta r\right>=0.07\pm0.04$~mm, which yields $k_w\approx 900~{\rm eV/mm}^2$.

\section{Vertical alignment}
\label{sec:VOP}

The strings that form inside DDC's in our experiments are highly aligned vertically. This is clearly seen in the strings' polar angle distribution, Fig.~\ref{angles} (top panel) and is also corroborated by large values (about $80-90$\%) of the vertical order parameter \cite{Killer:2011}. This is consistent with the results of dust dynamics simulations of Ref.~\cite{Ludwig:2012} for our estimated ion flow Mach number of $M=0.6-0.7$.

The polar angle probability densities happened to be similar in shape regardless of whether the DDC is clockwise or counterclockwise rotating, or even static. Such kind of universality is actually not surprising because the rotational degrees of freedom, in particular the polar angle distribution, must obey thermal statistics:
\begin{equation}
\label{eq_7}
P(\theta)=A\theta \exp\left(-\frac{\theta^2}{\left<\delta \theta^2\right>}\right), ~~~~\delta \theta=\theta-\left<\theta\right>,
\end{equation}
where $A$ is a coefficient and $\left<\delta \theta^2\right>$ is the mean squared variation. The value $\left<\delta \theta^2\right>\sim 0.01$ gives a fairly good fit for all cases shown in Fig.~\ref{angles}. By virtue of the fluctuation-dissipation theorem, it is straightforward to obtain the kinetic temperature $T$ of the particle chaotic vibrations in the DDC \footnote{Here, only as a kinetic average, since DDC's are not in equilibrium.}:
\begin{equation}
\label{eq_8}
T = \frac12 k_w \Delta_{\rm vert}^2\left<\delta \theta^2\right>\approx 3.2~eV.
\end{equation}
While being approximate, the relationship (\ref{eq_8}) provides a fairly good estimate of $T=(2/3)\left<mv^2/2\right>$, where the mean kinetic energy of particles $\left<mv^2/2\right>\approx 4$~eV was obtained directly by particle tracking technique.

\section{String-string interaction}
\label{subsec:InterStringInt}

The dynamics of paired particles in strings (as well as of isolated pairs) can be successfully used to study the in-string and string-string coupling constants. Evidently there are different time scales involved in this analysis. The short time scale evolution $\Delta t<0.1$~s can hardly be addressed properly in view of the strong pixel locking effect (for details, see e.g. Refs.~\cite{Feng:2007,Williams:2012}). The longest time scale of the order of a few rotation periods $\Delta t>20$~s is difficult to study since one needs to operate with enormously large amount of data. (In the given experiment we limited ourselves to $\Delta t=10$~s, which is approximately two thirds of the rotation period.) The remaining intermediate time scale $\Delta t\sim1$~s is the most important for our purposes. It allows one to study the evolution of the global structure of the cluster, which is determined predominantly by the mutually interacting strings.

The interaction between strings includes direct particle exchange \cite{Kroll:2010:1} and electrostatic interaction. From the spectrum of string deformations, the spring constant of the string-string interaction can be estimated. As a standard Fourier analysis revealed, the time scale of the string oscillations corresponds to frequencies of $0.5-3$~Hz, which are easy to study. Such a low frequency domain of string deformation can be successfully described assuming a dynamical balance between the friction force and the local force deforming the string. Assuming that the deformation is in the elasticity domain, with $k^*$ being the Hooke's spring constant, one can conclude that
\begin{equation}
\label{eq_3}
k^*\sim2\pi \gamma_{\rm eff} f_{\rm def}m,
\end{equation}
where $f_{\rm def}$ is a characteristic frequency of the low-frequency deformation. For $f_{\rm def}\sim 1-2$~Hz the relation (\ref{eq_3}) yields $k^*\sim600-1200~{\rm eV/mm}^2$. It is noteworthy that $k^*\sim k_w$.\footnote{The mutual particle-particle Yukawa interaction inside the string is $10-100$ times stronger \cite{Miloch:2010}.}


To conclude, 3D complex-plasma clusters driven externally by a rotating electric field were experimentally observed to be highly structured. They showed ion-flow aligned string lattices of a cylindrical symmetry. A fully 3D holographic particle tracking diagnostic allowed us to thoroughly explore the statistical as well as dynamical parameters of the clusters' string structure. Rotating electric field was shown to be a useful manipulation tool for 3D complex-plasma clusters.

\acknowledgments

The research leading to these results has received funding from the European Research Council under the European Union's Seventh Framework Programme (FP7/2007-2013) / ERC Grant agreement 267499, and has been financed by DFG under grant number SFB-TR24, project A3.
The authors would like to thank Dr. S\"utterlin and Dr. Konopka for fruitful discussions.

\bibliographystyle{eplbib}

\end{document}